\documentclass[aps,prl,twocolumn,groupedaddress]{revtex4-1}

\usepackage[dvipdfm]{graphicx}
\usepackage{subfigure}
\usepackage{here}
\usepackage{longtable}

\begin{document}

\title{Thermodynamic stability of Mg-Y-Zn ternary alloys
 through first-principles
}

\author{Ryohei Tanaka and Koretaka Yuge}

\affiliation{Department of materials science and engineering, Kyoto university, Kyoto, 606-8501, Japan}

\date{\today}

\begin{abstract}

In order to clarify thermodynamic stability of Mg-based long-period stacking ordered (LPSO) structure, we systematically study energetic preference for alloys on multiple stacking with different composition for random mixing of constituent elements, Mg, Y, and Zn based on special quasirandom structure (SQS).  
Through calculation of formation free energy of SQS, Mg-Y-Zn alloy exhibits phase separation into Mg- and Y-Zn rich phase, which is consistent with previous theoretical studies. Bulk modulus of SQSs for multiple compositions, stacking sequences, and atomic configurations ranges around 35 GPa, $i.e.$, they do not show significant dependence of Mg concentration, which therefore means that the effects of phonon do not play significant role on LPSO phase stability. Introducing stacking fault to hcp stacking gains "negative" energy, which indicates profound relationship between introducing stacking faults and the formation of long-period stacking ordering.

\end{abstract}


\maketitle

\section{1. Introduction}
\begin{figure}[]
\begin{center}
\includegraphics[width=8cm,clip]{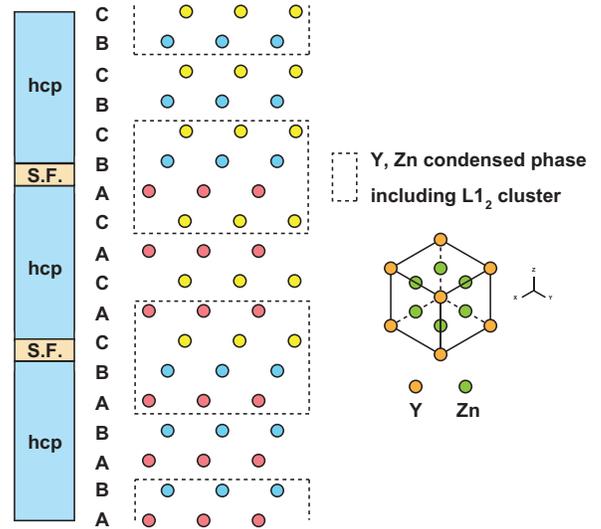}\\
\caption{Schematic illustration of stacking sequence of 18R LPSO structure. Stacking sequence is for hexagonal closed packed (hcp), which are denoted by A,B and C. "S.F." means stacking fault.}
\label{fig:18R}
\end{center}
\end{figure}

For its remarkable high tensile strength and ductility ~\cite{Kawamura}, Mg-based long-period stacking ordered (LPSO) structures are considered as light-weight structural alloy for next generation. In order to clarify relationship between thier formation process and resultant properties in terms of application for structural materials, considerable number of experimental as well as theoretical studies have been carried out. Previous theoretical studies mainly address formation process and thermodynamic stability of Mg-based LPSO alloy, including (i) the tendency of phase separation confirmed by Cluster variation method~\cite{Iikubo}, (ii) in-plane ordering of clusters consisted of Y, Zn substitutional atoms~\cite{Kimizuka}, and (iii) systematic understanding of energetic stability with respect to variety of substitutional atoms into Mg-based alloys ~\cite{Saal}. 
 Although these previous theoretical works partly clarify themodynamic stability of LPSO phases, they did not sufficiently discuss about (i) relative stability in terms of disordered phases or (ii) effect of lattice vibration on stability of LPSO. The former one is considered essentially important, since well-established ordering energy, determining the thermodynamic stability of ordered phase with respect to temperature, is typically reffered to the difference in mixing energy between ordered (here, LPSO)  and disordered phases. For the latter one, vibrational effects, their significant role on phase stability has been amply demonstrated for several binary alloys, such as Ref.~\cite{vandewalle}.
Fig.\ \ref{fig:18R} is a schematic illustration of stacking sequence of 18R LPSO structure, which includes Mg hcp stacking and Y, Zn substitutional atom concentrated phase including L1$_2$ cluster. To systematically evaluate thermodynamic stability of such structures, we need to consider energetics of multiple structures including different stacking sequence.
However, most calculations confine the structural model, so calculation that doesn't confine experimentally reported structures is highly required.
Furthermore, it is fundamentally important to assess thermodynamic stability of LPSO phase by competing disordered phase. In this study, based on DFT calculation, we systematically study thermodynamic stability of LPSO structure in terms of disordered phases.

\

\section{2. Methodology}

To comprehensively address energetics of multicomponent system with various composition, we need to calculate tremendous number of structures of the system and computational cost therefore become large. In this study, we employ special quasirandom structure (SQS) to assess thermodynamic stability of LPSO phase competing with disordered phase. SQS is a special microscopic state whose multiple correlation functions are numerically identical to those in perfect random structure, which therefore provides physical properties for perfect random alloy ~\cite{Zunger}. As following procedure, we calculated correlation function of SQS of ternary system.


Let us consider the system with $N$ lattice points for number of components, $R$. $\sigma _i$ is a variable which specifies the occupation of lattice point $i$ and $\vec{\sigma } = (\sigma _1,\sigma _2,\cdots,\sigma _N)$ can specify any atomic arrangement and we can represent structures in following equation, which is called correlation function;

\begin{equation}
\Phi_{\alpha}^M = \rho_{d_1}(\sigma _i)\rho_{d_2}(\sigma _j)\cdots\rho_{d_n}(\sigma _k).
\label{eq:SQS2}
\end{equation}

Here, $\rho(\sigma _i)$ is complete orthonormal basis function at lattice point $i$ and it is obtained by applying Gram-schmidt technique to the linearly independent polynominal set $(1,\sigma _i,\sigma _i^2,\cdots,\sigma_i^{R-1})$. $\alpha$ denotes a cluster included in the structure, $d_n$ is the index of basis function $\rho(\sigma _i)$ and $M$ is the set of index.
In the case of ternary alloy system, the occupation of lattice point by each element, Mg, Y, and Zn, is defined by $\sigma = +1,-1,0$ respectively, leading to the basis function of eq.(\ref{eq:basis_tri})~\cite{basis1}:\\

\begin{equation}
\rho_0 = 1, \rho_1 = \sqrt{\frac{3}{2}}\sigma, \rho_2 = -\sqrt{2}(1-\frac{3}{2}\sigma^2).
\label{eq:basis_tri}
\end{equation}

By averaging $\Phi_{\alpha}^M$ over equivalent clusters in lattice, correlation function of SQS in Mg-Y-Zn ternary system with compositions of each element, $x_{\rm Mg}$, $x_{\rm Y}$ and $x_{\rm Zn}$, $\overline{\Phi_{\alpha}^M}$, is represented by eq.(\ref{eq:SQS4}), 
 
\begin{equation}
\overline{\Phi_{\alpha}^M}= 2^{-\frac{n}{2}}\cdot3^{\frac{n-m}{2}}(x_{\rm Mg}-x_{\rm Y})^{n-m}(1-3x_{\rm Zn})^m ,
\label{eq:SQS4}
\end{equation}
where $n$ and $m$ denote dimension of clusters and the number of  $\rho_2$ respectively. The simulation was performed so that correlation functions come closer to ideal SQS correlation functions up to 4-th nearest neighbor pair ($i.e.$ 6 pair clusters on hcp)  finded by eq.(\ref{eq:SQS4}).
We optimized correration functions for each clusters by performing numerical simulation ~\cite{Yuge} and constructed SQSs. To evaluate the accuracy of obtained correlation functions of SQS, we compared the simulated values of correlation functions with standard deviation of them in configurational space. Then, most of the errors of correlation function are small enough compared with standard deviation. Based on these obtained structures, we calculate formation free energy, $F_{\rm form}$,  and bulk modulus, $B = - V \displaystyle{\frac{\partial^2E}{\partial V^2}}$ to evaluate thermodynamic stability of Mg-Y-Zn system that has multiple stacking sequence with various composition for random mixing.
We constructed structures with various composition on mutiple stacking sequence, whose composition is shown in Figure.\ \ref{fig:phasediagram}. The detail of calculation condition of structure is shown in Appendix.
 $F_{\rm form}$ is denoted by eq.(\ref{eq:form1}) and we define $U$ by eq.(\ref{eq:form2}),
 
\begin{equation}
 F_{\rm form} = U -TS
\label{eq:form1}
\end{equation}

\begin{equation}
 U = E({\rm Mg}_x{\rm Y}_y{\rm Zn}_z)-xE({\rm Mg})-yE({\rm Y})-zE({\rm Zn}),
\label{eq:form2}
\end{equation}
where $E(\gamma)$ is total energy of $\gamma$, and we evaluate configurational entropy, $S$, based on Bragg-Williams approximation.
In order to evaluate bulk modulus, we calculate total energies of structures, whose volume is expanded at a rate of $\pm$3\%, $\pm$6\%, $\pm$9\%. By using $B$ based on Debye-Gr\"{u}neisen approximation, Debye temperature, $\Theta_{\rm D}$, can be described as ~\cite{moruzzi}

\begin{equation}
\Theta_{\rm D} = 41.63\displaystyle\frac{(\displaystyle\frac{3}{4\pi}V_0)^{\frac{1}{3}}B_0^{\frac{1}{2}}}{M},
\label{eq:theta_d}
\end{equation}
where $V_0$ is atomic volume and $M$ denotes atomic mass.
Based on empirical Debye model, $F_{\rm vib}$ can be estimated by eq.(\ref{eq:fvib});

\begin{equation}
	F_{\rm vib} = \displaystyle\frac{9}{8}k\Theta_{\rm D} + kT\{3\ln(1-\exp(\frac{\Theta_{\rm D}}{T})-D(\frac{\Theta_{\rm D}}{T})\},
	\label{eq:fvib}
\end{equation}
where $D(x)$ is Debye function.

We estimated the effect of phonon on Mg-based alloy by calculating $\Delta F_{\rm vib}^{\rm form}$ and the vibrational free energy of formation with respect to the pure constituents can be estimated by eq.(\ref{eq:vibform})~\cite{ozolinz};

\begin{equation}
	\Delta F_{\rm vib}^{\rm form}(T) = F_{\rm vib}(T)- \sum_{i}x_iF_{\rm vib}^{i}(T),
	\label{eq:vibform}
\end{equation}
where $i$ denotes element and $x_i$ is a composition of $i$. 

In order to evaluate the effect of stacking faults on stability, we constructed a SQS which has stacking faults on hcp (stacking sequence is "ABABCABABCABABC"). Hereinafter, we call this structure "mixed". The effect of stacking difference is quantified by interfacial energy, which is defined by $E_{\rm I.F.}=\displaystyle\frac{E_{\rm mixed}-E_{\rm fcc,hcp}}{A}$, where $E_{\rm mixed}$ is total energy of structure including stacking faults and $E_{\rm fcc,hcp}$ is for structures on fcc and hcp stacking. $A$ denotes area of interface, which is 1.4$\times 10^2$ \AA$^2$ in this calculation. Additionally, to address stability of ordered phase competeing with disordered phase, we also calculated formation energy of structures including Y-Zn L1$_2$ cluster and estimated ordering temperature. 
We employ first-principles calculations using a DFT code, the Vienna Ab-initio Simulation Package (VASP)~\cite{vasp1}~\cite{vasp2}, to obtain the total energies for structures of Mg-Y-Zn alloys.  The calculation of total energy is carried out for the structures in Table.\ \ref{table:data_type}. All-electron Kohn-Sham equations are solved by employing the projector augmented-wave (PAW) method~\cite{paw1}~\cite{paw2}. We select generalized-gradient approximation of Perdew-Burke-Ernzerhof (GGA-PBE) ~\cite{ggapbe} form to the exchange-correlation functional. 
The plane-wave cutoff energy is set at 350 eV throughout the present calculations. Brillouin zone sampling is performed on the basis of the Monkhorst Pack scheme~\cite{monk}. K-point mesh is set 4$\times$4$\times$1 for structure, C3, and 4$\times$4$\times$4 for others and smearing parameter is 0.15 eV ~\cite{meth}.

\begin{figure}[H]
\begin{center}
\includegraphics[width=8.5cm,clip]{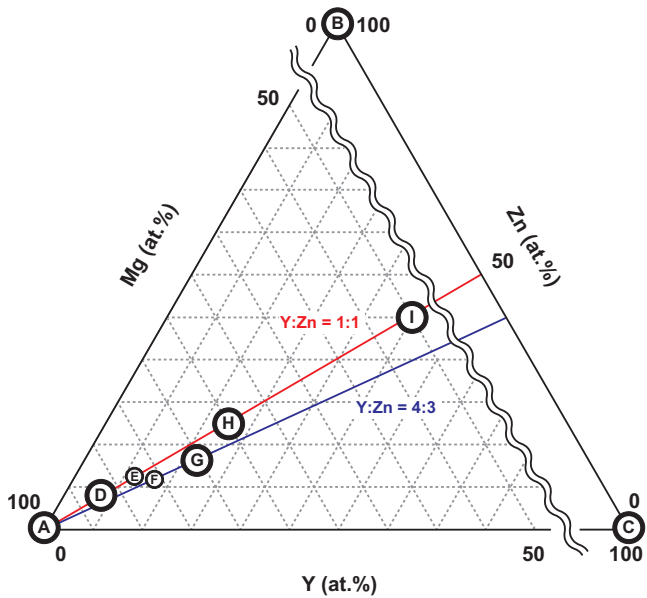}\\
\caption{Composition of structural model. Further information are denoted in Appendix.}
\label{fig:phasediagram}
\end{center}
\end{figure}\

\section{3. Results and Discussion}



\begin{figure}[]
	\begin{center}
		\includegraphics[width=9.5cm,clip]{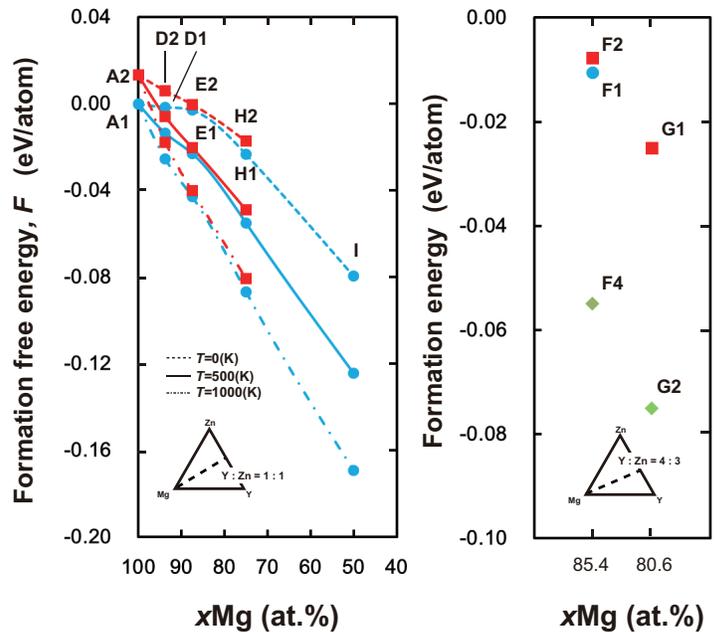}\\
		\caption{(left) Formation free energy of SQSs with multiple compositions on hcp and fcc stacking with respect to energies of structures on hcp. Curves are guide for the eyes, which shows tendency of phase separation. (right) Formation energy of structures including L1$_2$ cluster (F4, G2) and SQSs on hcp (F1) and fcc (F2, G1) stacking. The ratio of concentration of substitutional atom is denoted in each figure. Structure indices are denoted in Table.\ \ref{table:data_type}}
		\label{fig:Formation free energy}
	\end{center}
\end{figure}

First, we evaluated formation free energy of structures with multiple compositions on hcp and fcc stacking, which is shown in Fig.\ \ref{fig:Formation free energy}. At $T=500{\rm K}$, $F_{\rm form}$ of structures on hcp is negative at all through composition and possesses two extreme values. Through calculation of formation free energy of SQSs, Mg-Y-Zn alloy exhibits phase separation into Mg- and Y-Zn- rich phase and this result is consistent with previous research by Iikubo $et$ $al.$~\cite{Iikubo}, which suggests the validity of this simulation based on SQSs. Using these optimized SQSs, we proceed our discussion to calculate bulk modulus, ordering temperature, and the effect of stacking difference on phase stability. 

\begin{figure}[]
	\begin{center}
		\includegraphics[width=7cm,clip]{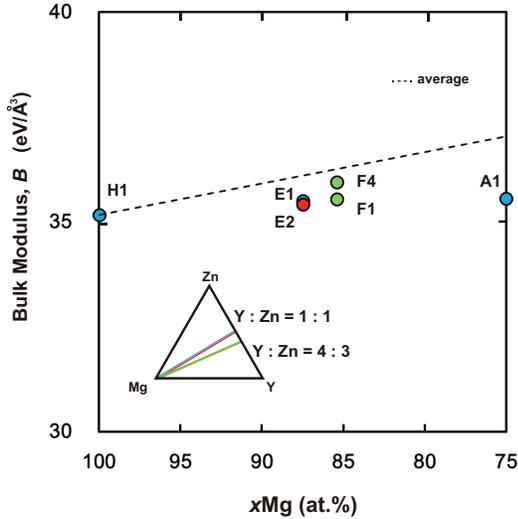}\\
		\caption{Bulk Modulus of SQSs and ordered structure. Dahsed line shows $B_{{\rm ave}} = xB_{{\rm Mg}}+yB_{{\rm Y}}+zB_{{\rm Zn}}$, where $B_A$ is bulk modulus of $A$ and $x$, $y$, $z$ denotes composition of each element. Colors correspond to the ratio of concentration of substitutional atom.}
		\label{fig:BulkModulus}
	\end{center}
\end{figure}

Next, in order to evaluate the effect of lattice vibration on phase stability of Mg-Y-Zn system, we calculated bulk modulus, which is shown in Figure.\ \ref{fig:BulkModulus}. Dashed line represents linear averaged bulk modulus, $B_{\rm ave}$. Bulk modulus for A1, E1 H1, F1 (on hcp stacking), E2 (on fcc stacking) and F4 (ordered structure) are about 35 GPa. The values of bulk modulus are smaller than $B_{\rm ave}$ through all compositions. Calculated $\Delta F_{\rm vib}^{\rm form}$ at $T=1000$K at ($x_{\rm Mg}$,$x_{\rm Y}$,$x_{\rm Zn}$)=(87.5,6.25,6.25) was found to be negligibly small within the calculated accuracy of $B$. Compared with avove results of $F_{\rm form}$ , $F_{\rm vib}$ does not have significant influence on $F_{\rm form}$ of Mg-Y-Zn alloys with multiple stackings and compositions. Additionally, bulk modulus does not show significant dependence of Mg concentration, which therefore means that the effects of phonon do not play essential role on LPSO phase stability. In this study, since effect of optical mode for multicomponent system is not considered, further study including optical mode is needed to quantitively clarify the effect of phonon for stability of Mg-Y-Zn alloys. Hereinafter, in evaluateing effects of ordering and stacking sequence difference on stability, we only consider the configurational effect.

\begin{figure}[]
  \begin{center}
    \includegraphics[width=7cm,clip]{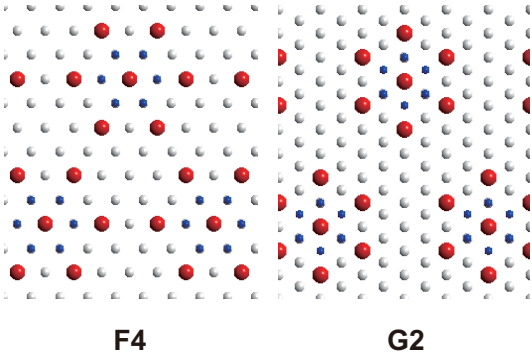}\\
    \caption{Schematic illustration of the atomic arrangements of structures including Y-Zn L1$_2$ clusters. Red and blue spheres represent Y and Zn atom respectively. F4 : Meta-stable. G2 : Stable.}
    \label{fig:L12clusterarrange}
  \end{center}
\end{figure}

 Then, we examine the effect of ordering on stability. We constructed structure including L1$_2$ cluster, F4 and G2, whose atomic arrangements are shown in Fig.\ \ref{fig:L12clusterarrange}. Formation energies of F4 and G2 are lower than SQS on hcp (F1) and fcc (G1,F2) stacking (right side of Fig.\ \ref{fig:Formation free energy}), which shows this system is stabilized by the effect of ordering. As we estimated ordering temperature, $T_{\rm_C}$, from F2 to F4 and from G1 to G2 was about 1050 K and 930 K respectively. This results shows that transition temperature depends on the arrangement of L1$_2$ cluster and  $T_{\rm_C}$ of G2 is close to experimental results by Okuda $et. al$. ~\cite{okuda} taking that transition temperature can be overestimated based on Bragg-Williams approximation into consideration. Moreover, this result indicates that Mg-Y-Zn system form ordered phase up to melting point: Mg-Y-Zn alloy can be regarded as intermetallic compounds. 

\begin{figure}[]
\begin{center}
\includegraphics[width=7cm,clip]{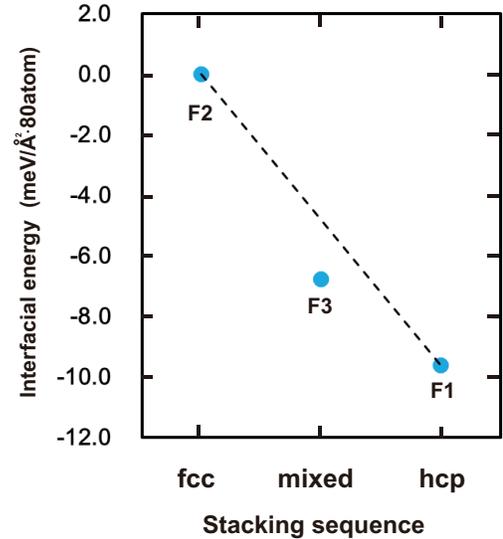}\\
\caption{The effect of stacking difference on interfacial energy of Mg-Y-Zn alloy on various stacking sequence}
\label{fig:stackingdifference}
\end{center}
\end{figure}

Finaly, we address the effect of stacking sequence on stability using interfacial energy. Fig.\ \ref{fig:stackingdifference} is interfacial energies of structures on different stacking (F1, F2, F3) with respect to the energy of structure on fcc stacking, F2. As shown in Fig.\ \ref{fig:stackingdifference}, interfacial energy of disordered phase that introduces stacking fault is lower than the energy for linear average of hcp and fcc stacking.
This result indicate that when stacking fault is introduced to Mg-rich hcp alloys to form "L1$_2$-like" ordering, corresponding interface between original hcp and formed fcc region gains "negative" energy, which is contrary to the conventional tendency that interface gains positive energy. This specific characteristics for interface energy in Mg-based allos would thus be one of the fundumental prerequisite to accelaration of forming LPSO phases, which should be further investigated in the future work.
 



\

\section{4. Conclusion}
In this study, we calculate preference of energetics of Mg-Y-Zn ternary alloy system in terms of disordered phase stability. Through calculation of formation free energy of SQSs, Mg-Y-Zn alloy exhibits phase separation into Mg- and Y-Zn- rich phase. Bulk modulus for SQSs and ordered structure ranges about 35 GPa and they do not show significant dependence of Mg concentration, which therefore means that the effect of phonon does not play essential role on LPSO phase stability within acoustic mode. Order-disorder transition temperature is estimated about 930 K and this results is colse to the melting point that is experimentaly reported and this suggests that Mg-Y-Zn LPSO alloy can be regarded as intermetallic compounds. The effect of stacking faults stabilized the LPSO phase and this results indicate that there remains profound relationship between introducing stacking faults and the formation of long-priod stacking ordering. 

\section{Acknowledgement}
\begin{acknowledgments}
This work is supported by a Grant-in-Aid for Scientific Research on Innovative Areas (26109710) from the Ministry of Education, Science, Sports and Culture of Japan.
\end{acknowledgments}



\appendix*
\setcounter{figure}{0}
\setcounter{table}{0}
\renewcommand{\thetable}{A\arabic{table}}
\renewcommand{\thefigure}{A\arabic{figure}}

¥¥

\section{Appendix}

In this study, we calculated free energy and bulk modulus of structures shown in following data.


\onecolumngrid

\begin{table}[hbtp]
  \caption{Composition, number of atoms, and stacking sequence of calculated structures}
  \label{table:data_type}
  \begin{center}
    \begin{tabular}{ccccccc}
      \hline
      index & Mg (at.\%)  & Y:Zn & number of atoms & stacking sequence & model & calculated bulk modulus (GPa)\\
      \hline \hline
      A1 &        &     & 96  & hcp   & pure Mg 	   & 35.2\\ 
      A2 &        &     & 96  & fcc   & pure Mg 	   &\\ 
      B1 &        &     & 96  & hcp   & pure Y 	       & 50.4 \\ 
      B2 &        &     & 96  & fcc   & pure Y 	       &\\ 
      C1 &        &     & 96  & hcp   & pure Zn		   & 34.9\\ 
      C2 &        &     & 96  & fcc   & pure Zn		   &\\ 
      D1 & 93.75  & 1:1 & 96  & hcp   & SQS            &\\
      D2 & 93.75  & 1:1 & 96  & fcc   & SQS            &\\
      E1 & 87.5   & 1:1 & 96  & hcp   & SQS            & 35.5\\
      E2 & 87.5   & 1:1 & 96  & fcc   & SQS			   & 35.4\\
	  F1 & 85.4   & 4:3 & 96  & hcp   & SQS 		   & 35.5\\
	  F2 & 85.4   & 4:3 & 96  & fcc   & SQS			   &\\
	  F3 & 85.4   & 4:3 & 240 & mixed & SQS			   &\\  
	  F4 & 85.4   & 4:3 & 96  & fcc   & Order (L1$_2$) & 35.9\\
	  G1 & 80.6   & 4:3 & 72  & fcc   & SQS &\\     
	  G2 & 80.6   & 4:3 & 72  & fcc   & Order (L1$_2$) &\\     
      H1 & 75     & 1:1 & 96  & hcp   & SQS			   & 35.6\\
      H2 & 75     & 1:1 & 96  & fcc   & SQS			   &\\
      I  & 50     & 1:1 & 96  & hcp   & SQS			   &\\
      \hline
    \end{tabular}
  \end{center}
\end{table}

\end{document}